\documentclass[aps,prd,twocolumn]{revtex4}
\usepackage{graphicx}
\usepackage{color}
\usepackage{amsmath}
\usepackage{amsfonts}
\usepackage{amssymb}

\begin{document}

\title{{\small{PHYSICAL REVIEW D {\bf{68}}, 123521 (2003)}\\
\vspace{0.3cm} Observational constraints on Chaplygin
quartessence: Background results}}
\author{Mart\'\i n Makler, S\'{e}rgio Quinet de Oliveira, and Ioav Waga}
\address{Universidade Federal do Rio de Janeiro, Instituto de F\'\i sica,
Caixa Postal 68528, CEP 21941-972 Rio de Janeiro, RJ, Brazil}
\date{Received 27 June 2003; published December 2003}
\begin{abstract}
We derive the constraints set by several experiments on the quartessence
Chaplygin model (QCM). In this scenario, a single fluid component drives the
Universe from a nonrelativistic matter-dominated phase to an accelerated
expansion phase behaving, first, like dark matter and in a more recent epoch
like dark energy. We consider current data from SNIa experiments, statistics
of gravitational lensing, FR IIb radio galaxies, and x-ray gas mass fraction
in galaxy clusters. We investigate the constraints from this data set on flat
Chaplygin quartessence cosmologies. The observables considered here are
dependent essentially on the background geometry, and not on the specific form
of the QCM fluctuations. We obtain the confidence region on the two parameters
of the model from a combined analysis of all the above tests. We find that the
best-fit occurs close to the $\Lambda$CDM limit ($\alpha=0$). The standard
Chaplygin quartessence ($\alpha=1$) is also allowed by the data, but only at
the $\sim2\sigma$ level.
\end{abstract}
\maketitle

\draft

\section{Introduction\label{intro}}

Over the past decade, a cosmological model consistent with most
astrophysical data available to date has emerged: a flat Universe
whose evolution is dominated by a repulsive cosmological term [or
dark energy (DE)] and pressureless cold dark matter (CDM). In
addition to observational evidence for this model, there are also
theoretical motivations for it: for example, inflation theory,
which generates a nearly flat space geometry and scale invariant
primordial perturbations. Another example regards the nature of
the two dark components. Particles predicted by extensions of the
standard model of interactions, such as the lightest supersymetric
(SUSY) particles \cite{susy} or the axion \cite{axion} are, for
instance, natural candidates for CDM. On the DE side, a slowly
rolling scalar field has been known to produce accelerated
expansion since the proposal of inflation. It is thus a well
motivated candidate for the cosmological term \cite{dynamical}. In
fact, the simplest and most popular candidate as the driving force
for the accelerated expansion is the cosmological constant
$\Lambda$. However, its tiny value inferred from present
observations creates a major puzzle. If $\Lambda$ is to be
considered as the sum of the vacuum energies of all fields, it is
hard to understand how they would cancel to one part in $10^{55}$
or 10$^{122}$ (for a review on the cosmological constant and DE,
see \cite{revde}).

In the standard cosmological model, dark matter and dark energy
are necessary to account for two seemingly independent
phenomena---clustering of matter and accelerated expansion. As we
pointed out, there are some theoretical hints about the nature of
these two components, but, in fact, at present, there is no
conclusive observational evidence that these phenomena are
produced by distinct components (see a first attempt to address
this question in Ref. \cite{sandvik02}). Therefore, instead of
using a theoretical bias towards dark energy plus CDM, one may
choose an alternative point of view and search for
phenomenological models that could still be consistent with
current observations, motivating theoretical investigations
\textit{a posteriori}. In addition to being potential candidates
for the dark sector, these models would allow us to test the
robustness of observational predictions regarding the
determination of the cosmological model.

From the point of view of simplicity, it would be interesting if
instead of two unknown components we could have a single one,
accounting for the phenomenology not associated with ordinary
matter. In this model, a single component would be responsible for
both clustering and accelerated expansion. Such a model is usually
referred to as UDM (unifying dark matter energy) or---since there
is only one dark component besides baryons, photons, and
neutrinos---as \textit{quartessence} \cite{makler03}.

The possibility of a unified description of DE-DM has given rise
to rather widespread interest recently. The two major candidates
for UDM are the quartessence Chaplygin fluid (QCM)
\cite{kamenshchik01,bilic02} and a quartessence tachyonic field
\cite{sen02}. Some confrontations of QCM against observational
data have been performed. It has been shown that, for a wide range
of parameters, this model is consistent with a number of tests of
the background metric
%\cite{fabris02,makler03,avelino03,colistete03,dev03,silva03}.
\cite{fabris02}\cite{makler03}\cite{avelino03,colistete03,dev03,silva03}.

Nevertheless, preliminary analyses of large-scale structure and CMB data favor
the $\Lambda$CDM limit of the QCM model \cite{sandvik02,amendola03}. However,
these perturbation analyses need further assumptions beyond the background
Chaplygin equation of state. For instance, they assume that there are no
viscous stresses in the perturbed fluid. In \cite{reis03} it was shown that,
if entropy perturbations are allowed, instabilities and oscillations, present
in the mass power spectrum in the adiabatic case, may be eliminated and, as a
consequence, the parameter space is enlarged.

The main goal of this paper is to set constraints on the Chaplygin
quartessence model and check whether it is consistent with present
cosmological data. We focus on observables that are dependent
essentially on the background geometry. We perform a combined
analysis of data, including SNIa experiments, gravitational
lensing statistics, FRIIb radio galaxies, and x-ray gas mass
fraction in galaxy clusters. Some of these tests were studied
previously within the QCM setting. We review these results with a
careful treatment and discuss the outcome of a combined analysis
of the data.

The paper is organized as follows. In the next section, we discuss the
phenomenological motivation for quartessence and present its realization
through a fluid with an exotic equation of state, focusing on the QCM.
Constraints from SNIa experiments are discussed in Sec. \ref{SNIa}. In Sec.
\ref{lensing}, we discuss the bounds on QCM parameters imposed by the
statistics of gravitational lenses. Limits from Fanaroff-Riley type IIb radio
galaxies are obtained in Sec. \ref{FRIIb} and from x-ray gas mass fraction in
galaxy clusters in Sec. \ref{xray}. We present a combined analysis of the
experiments and our concluding remarks in Sec. \ref{disc}.

\section{The Quartessence Chaplygin Model\label{4essence}}

As discussed in the previous section, the Universe is believed to be dominated
by two unknown components, generically denoted by dark energy and dark matter.
At the cosmological level, the direct detection of each of these two
components involves observations at different scales. Since it is not supposed
to cluster at small scales, the effect of dark energy can only be detected
over large distances, where the accelerated expansion is observed. On the
other hand, the CDM is detected by its local clustering through the motion of
visible matter or through the bending of light in gravitational lensing.

Within the standard lore of general relativity and the Friedmann--Lema\^{i}%
tre--Robertson--Walker model, for the Universe to undergo
acceleration its average density and pressure must obey
\begin{equation}
\left(  \rho+3p\right)  <0\,, \label{mgn}%
\end{equation}
so that there is an effective gravitational repulsive effect. On
the other hand, for the large-scale structures we see to have
formed today, the dark matter has to be nonrelativistic, i.e.,
\begin{equation}
\left|  p\right|  \ll\rho\,. \label{rmlp}%
\end{equation}
These two conditions are not in contradiction, since observations
in various scales probe different average densities. For example,
local motions in clusters of galaxies occur in regions hundreds of
times denser than the average density of the Universe. So, if the
pressure is negative [as needed from Eq. (\ref{mgn})] and the
ratio $\left|  p\right|  /\rho$ is a decreasing function of the
density, the two conditions can be made compatible. In this
picture, the quartessence would act as dark energy in very
low-density regions and as dark matter in higher-density regions.

A very simple equation of state that has the properties discussed above is the
so-called generalized Chaplygin gas \cite{kamenshchik01},\cite{makler01},
\cite{bilic02},\cite{bentoGCG},
\begin{equation}
p_{\text{Ch}}=-\frac{M^{4(\alpha+1)}}{\rho_{\text{Ch}}^{\alpha}}\,,
\label{eostoy}%
\end{equation}
where $M$ has the dimension of mass. Consider now the background geometry of
the Universe, which can be determined by assuming a homogeneous fluid whose
density and pressure are given by the averaged values of the present clumpy
matter distribution. In this case, the energy conservation equation
\begin{equation}
\dot{\rho}_{\text{Ch}}=-\left(  \rho_{\text{Ch}}+p_{\text{Ch}}\right)
3\frac{\dot{a}}{a} \label{rhop}%
\end{equation}
can be easily solved. The energy density of the Chaplygin fluid will be given
by
\begin{equation}
\rho_{\text{Ch}}=\rho_{\text{Ch0}}\left[  (1-A)\left(
\frac{a_{0}}{a}\right)
^{3(\alpha+1)}+A\right]  ^{1/(\alpha+1)}\,, \label{rhotoy}%
\end{equation}
where $a$ is the scale factor, $a_{0}$ is its present value, $A=(M^{4}%
/\rho_{ch0})^{(\alpha+1)}$, and the overdot in Eq. (\ref{rhop}) denotes the
derivative with respect to cosmic time. The equation-of-state parameter and
the adiabatic sound velocity, for the Chaplygin component, are given by
\begin{equation}
w_{\text{Ch}}(a)=\frac{p_{\text{Ch}}}{\rho_{\text{Ch}}}=-\frac{Aa^{3(\alpha
+1)}}{(1-A)+Aa^{3(\alpha+1)}} \label{wch}%
\end{equation}
and
\begin{equation}
c_{\text{sCh}}^{2}=\frac{\dot{p}_{\text{Ch}}}{\dot{\rho}_{\text{Ch}}}=-\alpha
w_{\text{Ch}}(a)\,. \label{csch}%
\end{equation}

In principle, the constant $M$ can take any positive value, but we
should have $M\sim10^{-3}$ eV in order to have negative pressure
at recent times \cite{makler03}. Note that, in QCM, we need
$A>1/(3-3\Omega_{b0})$ to have cosmic acceleration starting before
the present time ($q_{0}<0$). For instance, for
$\Omega_{\text{b}0}=0.04$, the above condition imposes the
constraint $A>0.347$. More generically, it is necessary that $A>0$
in order to have a positive cosmological constant at late times.
The adiabatic sound speed maximum value (which occurs in the
regions where $p_{\text{Ch}}\rightarrow -\rho_{\text{Ch}}$) is
given by $\sqrt{\alpha}$. Therefore, the parameter $\alpha$ is
restricted to the interval $\alpha\leq1$. The Chaplygin gas
$\alpha=1$ is the extreme case, where the sound velocity can be
nearly the speed of light. Another special case is $\alpha=0$,
which gives the equation of state
$p_{\text{Ch}}=-\rho_{\text{Ch0}}A$ and
$\rho_{\text{Ch}}=\rho_{\text{Ch0}}A+\rho_{\text{Ch0}}(1-A)(a_{0}/a)^{3}$,
and is, therefore, equivalent to a superposition of CDM and a
cosmological
constant. Note that, for $\alpha<-1$, we have at early times $w_{\text{Ch}%
}\sim-1$ and at late times $w_{\text{Ch}}\rightarrow0$. Thus, since we are
interested in the quartessence scenario, where at early times $w_{\text{Ch}%
}\sim0$ and at late times $w_{\text{Ch}}\sim-1$, we also impose
$\alpha>-1$. For $\rho_{\text{Ch}}$ to be well defined and/or
positive at early times, we also impose $A<1$. Therefore in QCM,
the parameter $\alpha$ is restricted to the interval
$-1<\alpha\leq1$, and $A$ is restricted to $0<A<1$. We can see
that, in fact, in QCM Eq. (\ref{rhotoy}) interpolates between dark
matter and dark energy as the average energy density of the
Universe changes. That is, when $a/a_{0}\ll1$, we have
$\rho_{\text{Ch}}\propto a^{-3}$ and the fluid
behaves as CDM. For late times, $a/a_{0}\gg1$, and we get $p_{\text{Ch}}%
=-\rho_{\text{Ch}}=-M^{4}=$ const as in the cosmological constant
case.

Equation (\ref{eostoy}) provides the simplest example of
quartessence. It has naturally a single-dimensional constant,
since it is given by a power law, and allows us to solve the
background energy conservation equation analytically.
Besides, it has only two free parameters. However, many other \textit{Ans\"{a}%
tze} for equations of state (EOS) satisfying these criteria can be found.

So far we have introduced the QCM from a purely phenomenological
point of view. The EOS in Eq. (\ref{eostoy}) was chosen as a
simple toy model that could allow dark energy/dark matter
unification. A completely independent point of view, motivated by
brane dynamics, led to the same EOS. Kamenshchik \textit{et al.}
\cite{kamenshchik01} proposed EOS in Eq. (\ref{eostoy}), with
$\alpha=1$ (the standard Chaplygin case) as a dark energy
candidate. Only after was it realized that such fluid could
naturally lead to dark energy/dark matter unification
\cite{bilic02}. Some possible motivations for this scenario from
the field theory point of view are discussed in Refs.
\cite{kamenshchik01,bilic02,bentoGCG}. The Chaplygin gas appears
as an effective fluid associated with the parametrization
invariant Nambu-Goto $d$-brane action in a ($d+1,1$) spacetime,
and it can also be derived from a Born-Infeld Lagrangian
\cite{Jackiw99,kamenshchik01}. The generalized Chaplygin EOS can
be obtained from a complex scalar field Lagrangian with
appropriate potential \cite{bilic02,bentoGCG}. The relation
between a Chaplygin-like gas and the tachyonic scalar field was
investigated in \cite{Benaoum}.

In the rest of this paper, we will consider only the case of the Chaplygin
fluid as quartessence. We will place constraints on the QCM parameters from
several cosmological tests which are dependent essentially on the background
geometry. We emphasize that the resulting constraints depend on our choice of
priors. For example, in Chaplygin quintessence it is commonly assumed that
$\Omega_{\text{CDM}}\sim0.3$, while here we assume $\Omega_{\text{CDM}}=0$.
Therefore, the limits on the parameters $\alpha$ and $A$ will be different.

A fundamental quantity related to the observables considered here is the
distance-redshift relation, given by
\begin{equation}
r\left(  z\right)  =\int_{0}^{z}\left[  H\left(  z^{\prime}\right)  \right]
^{-1}dz^{\prime}\,, \label{rz}%
\end{equation}
where $H$ is the Hubble parameter. In the QCM case (neglecting the radiation
component), we have
\begin{align}
H\left(  z\right)   &  =H_{0}\left(  \left\{  \Omega_{b}\left(  1+z\right)
^{3}\right.  \right. \nonumber\\
&  +\left.  \left.  \Omega_{\text{Ch}}\left[  A+(1-A)\left(  1+z\right)
^{3\left(  1+\alpha\right)  }\right]  ^{1/\left(  1+\alpha\right)  }\right\}
\right)  ^{1/2}\,, \label{EzTot}%
\end{align}
where $H_{0}$ is the Hubble constant, $z=a_{0}/a-1$ is the redshift, and
$\Omega_{\text{Ch}}$, $\Omega_{b}$ are the density parameters of the Chaplygin
fluid and baryons, respectively.

Notice that following the idea of unification, we will not include
an additional dark matter component. Thus, in Eq. (\ref{EzTot})
only the baryonic matter scales as $(1+z)^{3}$. Just for $a\ll
a_{0}$ does the Chaplygin component scale as CDM. In this case, we
have an effective matter density parameter$^{1}$
\footnotetext[1]{This is why we have used the parametrization
$\Omega_{M}^{\star}:=(1-A)$ in Ref. \cite{makler03} that
corresponds to the effective matter density of the generalized
Chaplygin gas for
$\Omega_{\text{Ch}}=1$ and $\alpha=0$.}%

\begin{equation}
\Omega_{m}^{\text{eff}}=\Omega_{\text{Ch}}\left(  1-A\right)  ^{1/\left(
1+\alpha\right)  }+\Omega_{b}\,. \label{OmMeff}%
\end{equation}

Notice that, after radiation domination, in this scenario the
Universe is always quartessence-dominated, thus avoiding the ``why
now'' problem. However, there is still some fine-tuning to
determine whether the average equation of state turns from
``matter like'' to ``energy like'' (i.e, when the quartessence
pressure begins to play an important role \cite{makler03}).

Because after the decoupling epoch the QCM has negligible
pressure, it will cluster as ordinary CDM. In this model, only the
unclustered part would lately have appreciable negative pressure.

Since observations of anisotropies in the CMB indicate that the Universe is
nearly flat, and since the inflation paradigm predicts a flat geometry, we
restrict the following discussions to the zero curvature case, such that
$\Omega_{\text{Ch}}+\Omega_{b}=1$.

The baryon density and Hubble parameters can be determined
independently of the quartessence model. The observed abundances
of light elements together with primordial nucleosynthesis give
$\Omega_{b}h^{2}=0.0214\pm0.0018$ \cite{BBN,DH}. The Hubble Space
Telescope (HST) key project result is $h=0.72\pm0.08$
\cite{freedman}, where $H_{0}:=100h\,$Km/sec/Mpc. These bounds on
$\Omega_{b}h^{2}$ and $h$ are in agreement with a fit of purely
Wilkinson Microwave Anisotropy Probe (WMAP) and other Cosmic
Microwave Background (CMB) data \cite{WMAPcospar}. Thus,
throughout most of this paper we fix the Hubble and the baryon
density parameters at $h=0.72$ and $\Omega_{b}=0.041$. With
$\Omega_{b}$ and $h$ being determined by independent measurements,
in the following sections we will investigate the constraints on
the parameters $A$ and $\alpha$ from several experiments, assuming
the above-stated priors.

\section{Type Ia Supernovae Experiments\label{SNIa}}

Supernovae constraints on the Chaplygin models have been analyzed by many
authors \cite{fabris02},\cite{makler03,avelino03,amendola03}, using different
priors and simplifications. Here we repeat the analysis presented in
\cite{makler03}, but we shall not impose any prior on the parameter $A$.

The luminosity distance of a light source is defined in such a way as to
generalize to an expanding and curved space the inverse-square law of
brightness valid in a static Euclidean space,%

\begin{equation}
d_{L}=\left(  \frac{L}{4\pi\mathcal{F}}\right)  ^{1/2}=(1+z)r(z)\,. \label{L}%
\end{equation}
In Eq. (\ref{L}), ${L}$ is the absolute luminosity, $\mathcal{F}$
is the measured flux, and $r(z)$ is given by Eq. (\ref{rz}).

For a source of absolute magnitude $M$, the apparent bolometric magnitude
$m(z)$ can be expressed as
\begin{equation}
m(z)=\mathcal{M}+5\log D_{L}\,, \label{appmag}%
\end{equation}
where $D_{L}=D_{L}(z,\alpha,A)$ is the luminosity distance in units of
$H_{0}^{-1}$, and
\begin{equation}
\mathcal{M}=M-5\log H_{0}+25
\end{equation}
is the ``zero point'' magnitude (or Hubble intercept magnitude).

We follow the Bayesian approach of Drell, Loredo and Wasserman \cite{drell}
and consider the data of fit C of Perlmutter\textit{\ et al.}
\cite{perlmutter} with 16 low-redshift and 38 high-redshift supernovae. In our
analysis, we use the following marginal likelihood:
\begin{equation}
\mathcal{L}(\alpha,A)=\frac{s\sqrt{2\pi}}{\Delta\eta}e^{-q/2}\,.
\end{equation}
Here
\begin{align}
q(\alpha,A)  &  =\sum\limits_{i=1}^{16}\frac{(-5\text{log}D_{L}+\nu%
+m_{\text{B}i}^{\text{corr}})^{2}}{\sigma_{\text{low,}i}^{2}}\nonumber\\
&  +\sum\limits_{i=1}^{38}\frac{\left(  -5\text{log}D_{L}+\nu+m_{\text{B}%
i}^{\text{eff}}\right)  ^{2}}{\sigma_{\text{high},i}^{2}}\,,
\end{align}
where
\begin{align}
\nu(\alpha,A)  &  =s^{2}\,\left(  \sum\limits_{i=1}^{16}\frac{5\text{log}%
D_{L}(z_{i},\alpha,A)-m_{\text{Bi}}^{\text{corr}}}{\sigma_{\text{low,}i}^{2}%
}\right. \nonumber\\
&  +\left.
\sum\limits_{i=1}^{38}\frac{5\text{log}D_{L}(z_{i},\alpha
,A)-m_{\text{B}i}^{\text{eff}}}{\sigma_{\text{high,}i}^{2}}\right)
\,,
\end{align}%

\begin{equation}
s^{2}=\left(  \sum\limits_{i=1}^{16}\frac{1}{\sigma_{\text{low,}i}^{2}}%
+\sum\limits_{i=1}^{38}\frac{1}{\sigma_{\text{high,}i}^{2}}\right)
^{-1}\,,
\end{equation}%

\begin{equation}
\sigma_{\text{low,}i}^{2}=\sigma_{m_{B,i}^{\text{corr}}}^{2}+\left(
\frac{5\;\text{log\thinspace}e}{z_{i}}\sigma_{z_{i}}\right)  ^{2}%
\end{equation}
and
\begin{equation}
\sigma_{\text{high,}i}^{2}=\sigma_{m_{B,i}^{\text{eff}}}^{2}+\left(
\frac{5\;\text{log\thinspace}e}{z_{i}}\sigma_{z_{i}}\right)
^{2}\,.
\end{equation}
The quantities $m_{B}^{\text{corr}}$, $m_{B}^{\text{eff}}$, $\sigma
_{m_{B}^{\text{corr}}}$, $\sigma_{m_{B}^{\text{eff}}}$ and $\sigma_{z}$ are
given in Tables 1 and 2 of Perlmutter\textit{\ et al.} \cite{perlmutter}.

The results of our analysis for the QCM, marginalizing over the
intercept, are displayed in Fig. $1$. In this figure, we show
$68.3$ and $95.4$ confidence level contours in the $(\alpha,$ $A)$
plane. As we had observed in \cite{makler03}, current SNIa data
constrain $A$ to the range $0.4\lesssim A\lesssim0.9$, but do not
strongly constrain the parameter $\alpha$ in the considered range.

\begin{figure}[ptb]
\centering\hspace*{+0.3in} \includegraphics[height= 7.5
cm,width=8.0cm]{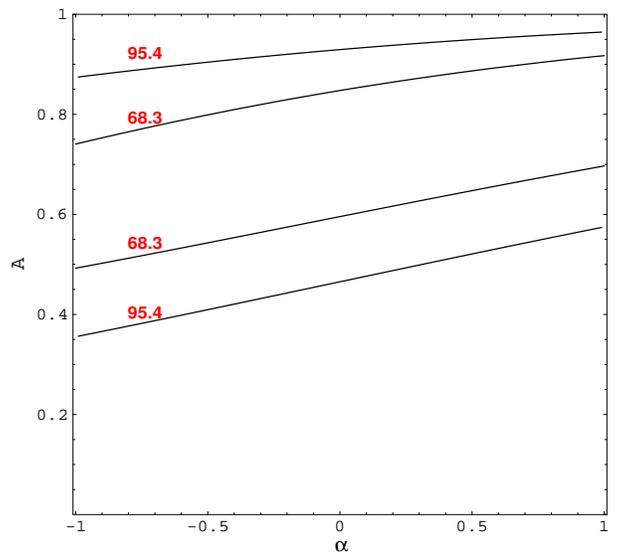} \caption{Constant confidence contours
($68.3\%$ and $95.4\%$) in the ($\alpha,A$) plane of the QCM
parameters allowed by SNIa data, as described in the text.}
\end{figure}

\section{Gravitational Lensing Statistics\label{lensing}}

The statistics of gravitational lensing \cite{tog} is one of the
most traditional and widely used methods to constrain cosmological
models, especially those with a cosmological constant. Since the
publication of the first works on lensing statistics
\cite{fukugita90,fukugita92,turn} almost 15 years ago, several
studies constraining $\Lambda$CDM models \cite{krauss,maoz,koch93}
and some of its variants \cite{ratra92,bloom,cooray} appeared in
the literature. Although not without controversy, before 1998 the
general belief was that models with $\Omega_{m0}\sim1$ and
$\Omega_{\Lambda 0}\sim0$ are preferred by lensing calculations.
These results started to be challenged, a few years ago, when the
high redshift supernovae observations, in combination with CMB
measurements, consistently indicated a nearly flat, low-density,
and accelerated Universe. In fact, lensing estimates were not in
strong conflict with the supernovae observations. However, it
cannot be said also that they were in comfortable agreement. There
was room in the parameter space for concordance, mainly if
$\Lambda$ is dynamical, but the best-fit regions for each one of
the tests were not in good agreement.

With few exceptions, in the past most lensing analyses were based
on optically selected quasars. One important concern in lensing
investigations with optical sources is extinction in the lensing
galaxies. The presence of dust in the lens galaxy could make the
quasar images fainter and, as a consequence, difficult to detect.
Radio selected surveys are immune to extinction. Besides reducing
the total number of multiply imaged systems in an optically
selected sample, extinction also affects its angular separation
distribution and this is very relevant in lensing analyses. In
most studies with optically selected sources, extinction has been
neglected. The motivation for this assumption is justified by the
fact that early-type galaxies that dominate the lensing statistics
are believed to have little dust at the present epoch. However,
the possibility that the existence of dust in early-type galaxies
at high redshifts ($z>0.5$) could reconcile lensing calculations
with models with high
$\Lambda$ was suggested 10 years ago by Fukugita and Peebles \cite{fukugita93}%
. Malhotra, Rhoads, and Turner \cite{malhotra} explored this
possibility further and found evidence for it. They estimated a
mean extinction of $\Delta m=2\pm1$ magnitudes. By observing that
statistical lensing analyses based on optical and radio
observations were not in accordance, Falco \textit{et al.}
\cite{fal98} suggested that they can be reconciled if the
existence of dust in E/S0 galaxies is considered. However, they
estimated that the required mean extinction is considerably lower
than that estimated in \cite{malhotra}. They obtained $\Delta
m=0.6\pm0.4$ mag. In a subsequent work, Falco \textit{et al.}
\cite{fal99} concluded that a substantial correction for
extinction is necessary in any cosmological estimate using the
statistics of lensed quasars. Their directly measured extinction
distribution was consistent with statistical estimates from
comparison of radio-selected and optically selected lens surveys
\cite{fal98}.

Another important source of uncertainty in gravitational lensing
investigations is the velocity dispersion of early-type galaxies
($\sigma_{\ast}^{(e)}$). The probability of lensing, or the
optical depth ($\tau$), depends on the fourth power of
$\sigma_{\ast}^{(e)}$ and, hence, is very sensitive to this
quantity. In \cite{koch96}, Kochanek advocated a high value for
the velocity dispersion, namely $\sigma_{\ast}^{(e)}\simeq225$
Km/sec. Lensing analyses that use relatively high values of
$\sigma_{\ast }^{(e)}$ predict low values for $\Lambda$. The
reason for this is simple: high $\Lambda$ values imply larger
distances and, as a consequence, also larger probability for
lensing. In order to have $\tau$ in high-$\Lambda$ models in
accordance with observations, one possibility is to ``compensate''
its increase due to $\Lambda$ by a reduction in
$\sigma_{\ast}^{(e)}$. However, this is not so easy. The angular
image separation, another observable in lensing statistics,
depends on the second power of $\sigma_{\ast}^{(e)}$. Therefore, a
too low value of $\sigma_{\ast}^{(e)}$ may not fit the image
separation distribution observed in optically selected lens
surveys. In fact, the main support for a high
$\sigma_{\ast}^{(e)}$ value comes from the angular image
separation distribution observed in these surveys. The observed
value in the sample used in \cite{koch96} is $\left\langle
\Delta\theta\right\rangle \simeq1.6$ $\operatorname{arcsec}$. The
sample of radio selected multiply imaged sources used by Falco
\textit{et al.} \cite{fal98} in their comparison with the
optically selected ones also has a similar high $\left\langle
\Delta\theta\right\rangle $.

The situation described above changed considerably with the completion of the
Cosmic Lens All Sky Survey (CLASS) sample of radio sources
\cite{mayers,browne}. Recently, Chae and Chae \textit{et al.} \cite{chae}
performed a statistical lensing analysis with CLASS data. Under a well-defined
selection criterion, they selected a total of 8958 radio sources from the
CLASS. From these, a total of 13 systems have multiple images, giving a
lensing rate of $\simeq1/689$. An important difference between the new data
and previous ones is that in this sample the mean angular image separation is
considerably smaller, $\left\langle \Delta\theta\right\rangle \simeq1.2$
$\operatorname{arcsec}$. By assuming a singular isothermal ellipsoid as the
lens model and a ``steep'' faint-end slope of early-type galaxies luminosity
function ($\alpha^{(e)}=-1$), they determined the velocity dispersion to be
$\sigma_{\ast}^{(e)}\simeq198$ Km/sec. This value is significantly smaller
than the best-fit value obtained in \cite{koch96} from optically selected
surveys. By assuming a flat $\Lambda$CDM model, their likelihood analysis
gives ${\Omega_{m0}=0.31}_{-0.14}^{+0.27}$ at the $68\%$ confidence level.
They also obtained results without fixing the curvature but fixing the
equation of state of the dark energy ($w$) to be $w=-1$, and/or fixing the
curvature to be zero and taking $w$ constant but not necessarily equal to
$-1$. Their results indicate that, although lensing statistics is less
restrictive than the magnitude-redshift test with type Ia supernovae, the two
tests are in very good agreement.

Although lensing analyses with radio sources are more reliable, in this work
we still use optically selected quasars. We leave for future work the
inclusion of radio sources in our analysis.

In the following, we briefly outline our main assumptions for the
lensing analysis using highly luminous quasars \cite{koch93,wm99}.
We consider data from the HST Snapshot survey [498 highly luminous
quasars (HLQ)], the Crampton survey (43 HLQ), the Yee survey (37
HLQ), the ESO/Liege survey (61 HLQ), the HST GO observations (17
HLQ), the CFA survey (102 HLQ), and the NOT survey (104 HLQ)
\cite{mao}. We consider a total of 862 ($z>1$) highly luminous
optical quasars plus five lenses. The lens galaxies are modeled as
singular isothermal spheres (SIS), and we consider lensing only by
early-type galaxies, since they dominate the lens population. We
assume a conserved comoving number density of early-type galaxies
\cite{peebles02}, $n_{e}=n_{0}(1+z)^{3}$, and a
Schechter form \cite{sch76} for the early-type galaxy population, $n_{0}%
=\int_{0}^{\infty}n_{\ast}\left(  L/L^{\ast}\right)  ^{\alpha_{\ast}}%
\exp\left(  -L/L^{\ast}\right)  dL/L^{\ast},$ with $n_{\ast}=0.64\pm
0.19\;h^{3}10^{-2}$ $\mbox{Mpc}^{-3}$ and $\alpha_{\ast}=-1.0\pm0.09$
\cite{marzke98,chae}. We assume that the luminosity satisfies the
Faber-Jackson relation \cite{fab76}, ${L}/{L^{\ast}}=({\sigma}/{\sigma_{\ast}%
})^{\gamma}$, with $\gamma=4$ and adopt $\sigma_{\ast}=198$
Km/sec. We emphasize that, even if we consider velocity dispersion
of early-type galaxies as low as $\sigma_{\ast}\simeq190$ Km/sec,
\textit{extinction has to be considered} in order to reconcile the
statistical lensing analysis using optically selected sources with
those based on CLASS radio-selected sources. Further, if we use
$n_{\ast}=0.99\pm0.05h^{3}10^{-2}$Mpc$^{-3}$ and
$\alpha_{\ast}=-0.54\pm0.02$, as obtained by Madgwick \textit{et
al.} \cite{madgwick02}, it would be necessary to consider a higher
amount of extinction. Previous analyses with optically selected
quasars that use high values for $\sigma_{\ast}$ and do not
consider extinction are not consistent.

For SIS, the total optical depth ($\tau$) can be expressed analytically,
$\tau(z_{S})=F/30\left[  d_{A}(1+z_{S})\right]  ^{3}(cH_{0}^{-1})^{-3},$ where
$z_{s}$ is the source redshift, $d_{A}=r(z)/(1+z)$ is its angular diameter
distance, and $F=16\pi^{3}n_{\ast}(cH_{0}^{-1})^{3}(\sigma_{\ast}/c)^{4}%
\Gamma(1+\alpha_{\ast}+4/\gamma)\simeq0.016$ measures the effectiveness of the
lens in producing multiple images \cite{tog}. We correct the optical depth for
the effects of magnification bias and include the selection function due to
finite angular resolution and dynamic range \cite{koch93,koch96}. We assume a
mean extinction of $\Delta m=0.6$ mag; this makes the lensing statistics for
optically selected quasars consistent with the results of \cite{chae}.

In Fig. $2$, we show contours of constant likelihood ($95.4\%$ and $68.3\%$)
arising from lensing statistics for the quartessence Chaplygin model. It is
clear from the figure that lensing statistics weakly constrains these models.
Only models with $A\gtrsim1$ are excluded by the lensing data.

\begin{figure}[ptb]
\centering\hspace*{-0.8in} \includegraphics[height= 7.5
cm,width=8.0cm]{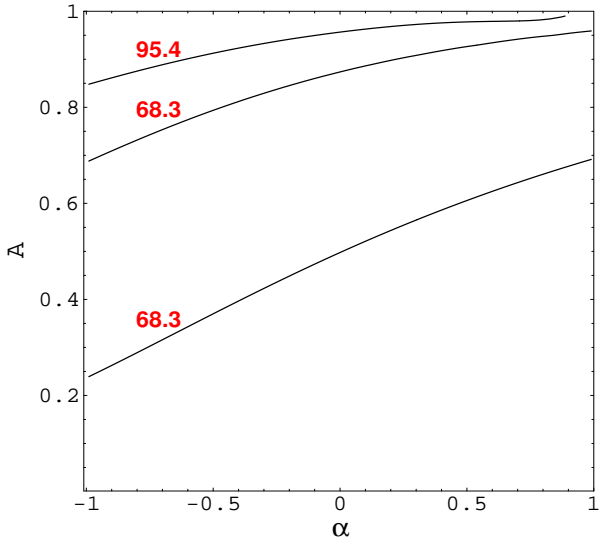}
\caption{Confidence contours in the
($\alpha,A$) plane from the statistics of
optically selected gravitational lenses.}%
\end{figure}

\section{Fanaroff-Riley Type IIb Radio Galaxies\label{FRIIb}}

In addition to gravitational lensing statistics and the SNIa
magnitude-redshift test, another useful method to constrain cosmological
parameters is the classical angular size-redshift test. In this paper, we
shall be concerned with the Fanaroff-Riley type IIb (FR IIb) radio galaxy
\cite{fanaroff74} version of this test as proposed in \cite{daly94} (see also
\cite{guerra98,guerra00,daly02,podariu03}). This test consists of a comparison
of two independent measures of the average size of the lobe-lobe separation of
FR IIb sources, namely the mean size $\left\langle D\right\rangle $ of the
full population of radio galaxies at similar redshift and the source average
(over its entire life) size $D_{\ast}$, which is determined via a physical
model that describes the evolution of the source. The basic idea is that
$\left\langle D\right\rangle $ must track the value of $D_{\ast}$, such that
the ratio $R_{\ast}=\left\langle D\right\rangle /D_{\ast}$ is independent of
redshift. It can be shown that $R_{\ast}\propto r^{(2\beta/3+3/7)}$, where $r$
is the comoving distance, and $\beta$ is a parameter to be determined
\cite{guerra00}. To determine the confidence region of the parameters of the
model, we use the following $\chi^{2}$ function:
\begin{equation}
\chi^{2}=\sum\limits_{i=1}^{20}\frac{\left[  R_{\ast},_{i}-c(r,_{i}%
/r)^{(2\beta/3+3/7)}\right]  ^{2}}{\sigma_{i}^{2}}\,,
\end{equation}
where $r,_{i}=r(z_{i},\Omega_{m0}=0.1,\Omega_{\Lambda0}=0)$, $\sigma_{i}$ is
the combination of the errors in $\left\langle D\right\rangle $ and $D_{\ast}%
$, and $c$ is a parameter that minimizes the $\chi^{2}$ for fixed
values of the cosmological parameters. In our computation, we
marginalize over $\beta$ assuming that it is Gaussian distributed
such that $\beta=1.75\pm0.25$ \cite{guerra00}.

\begin{figure}[ptb]
\centering\hspace*{0.in} \includegraphics[height= 7.5
cm,width=8.0cm]{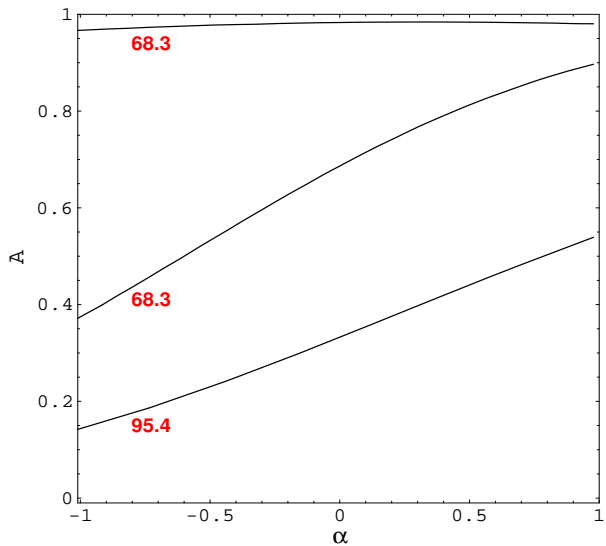} \caption{Confidence contours in the
($\alpha,A$) plane allowed by a set of FR IIb radio galaxies.}
\label{fig3}
\end{figure}

In Fig.~\ref{fig3}, we show contours of constant likelihood
($95.4\%$ and $68.3\%$) arising from the radio galaxies test for
the quartessence Chaplygin model. We can see that the FR IIb radio
galaxies test also does not strongly constrain the QCM models.
Only models with $A\lesssim0.2$ are excluded by the data at the
$95.4\%$ confidence level. However, as we shall see in
Sec.\ref{disc}, when we combine this test with the strong lensing
test, we get results similar to those obtained from SNIa.

\section{X-ray gas mass fraction in galaxy clusters\label{xray}}

In recent years, considerable efforts have been devoted to determining the
matter content of clusters of galaxies. Clusters of galaxies are the most
recent large-scale structures formed and are also the largest gravitationally
bound systems known. Therefore, the determination of their matter content is
quite important because cluster properties should approach those of the
Universe as a whole. A powerful method based on this idea is to measure the
baryon mass fraction $\Omega_{b}/\Omega_{m}$ in rich clusters. By combining
this ratio with $\Omega_{b}$ determinations from primordial nucleosynthesis,
strong constraints on $\Omega_{m}$ can be placed \cite{white}. As we shall
show, this is especially interesting in quartessence models such as QCM; since
the dark sector is unified in these models, any strong constraint on effective
dark matter translates directly into a strong constraint on effective dark energy.

Here we use the method and data of Allen, Schmidt, and Fabian \cite{allen02}
and Allen \textit{et al.} \cite{allen03}. These authors extracted from Chandra
observations the x-ray gas mass fraction $f_{\text{gas}}$ of nine massive,
dynamically relaxed galaxy clusters, with redshifts in the range
$0.08<z<0.47$, and that have converging $f_{\text{gas}}$ within a radius
$r_{2500}$ (radius encompassing a region with mean mass density $2500$ times
the critical density of the Universe at the cluster redshift).

To determine the confidence region of the parameters of the model, we use the
following $\chi^{2}$ function in our computation:
\begin{equation}
\chi^{2}=\sum\limits_{i=1}^{9}\frac{\left[  f_{\text{gas}}^{\text{mod}}%
(z_{i})-f_{\text{gas},i}\right]
^{2}}{\sigma_{f_{\text{gas},i}}^{2}}\,,
\label{chi2cluster}%
\end{equation}
where $z_{i}$, $f_{\text{gas},i}$, and $\sigma_{f_{\text{gas},i}}$
are, respectively, the redshifts, the SCDM ($h=0.5$) best-fitting
values, and the symmetric root-mean-square errors for the nine
clusters as given in \cite{allen02} and \cite{allen03}. In
Eq.~(\ref{chi2cluster}), $f_{\text{gas}}^{\text{mod}}$ is the
model function \cite{allen02}
\begin{equation}
f_{\text{gas}}^{\text{mod}}(z)=\frac{b\Omega_{b}}{(1+0.19\sqrt{h})\Omega
_{m}^{\text{eff}}}\left(  \frac{h}{0.5}\frac{d_{A}^{\text{EdS}}}{d_{A}%
^{\alpha,A}}\right)  ^{3/2}\,. \label{fgasmod}%
\end{equation}
Here, $d_{A}$ is the angular diameter distance to the cluster,
$\Omega _{m}^{\text{eff}}$, given by Eq. (\ref{OmMeff}), is the
effective matter density parameter, and $b$ is a bias factor that
takes into account the fact that the baryon fraction in clusters
could be lower than for the Universe as a whole. In our
computations, we marginalize over the bias factor assuming that it
is Gaussian-distributed with $b=0.93\pm0.05$ as suggested by
gas-dynamical simulations$^{2}$ \cite{bialek01,allen03}.
\footnotetext[2]{We also considered $b=0.824\pm0.033$ for the bias
factor (see \cite{eke98,allen03b}). In this case, the resulting
contours are slightly thinner than those in Fig. $4$, and the
best-fit value for the parameters occurs at $\alpha\simeq-0.12$
and $A\simeq0.72$.}

In Fig. $4,$ we show the $68.3\%$ and $95.4\%$ confidence contours on the
parameters $\alpha$ and $A$ determined from the Chandra data. The best-fit
value, the solid dot in the figure, is located at $\alpha\simeq0$ and
$A\simeq0.73$ and, as discussed in Sec.\ref{4essence}, corresponds to the
$\Lambda$CDM limit of QCM. It is clear from the figure that this test is much
more restrictive than the others discussed previously.

\begin{figure}[htb]
\centering\hspace*{-0.8in} \includegraphics[height= 7.5
cm,width=8.0cm]{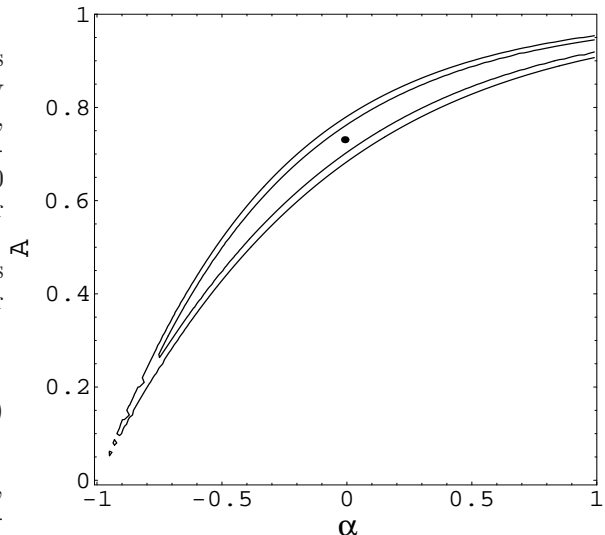} \caption{Constant confidence contours
($68.3\%$ and $95.4\%$) in the ($\alpha,A$) plane determined from
the x-ray gas mass fraction in nine galaxy clusters from Chandra
data. The best-fit value is indicated by a dot at the
center of the contours.}%
\end{figure}

\begin{figure}[htb]
\centering\hspace*{-0.8in} \includegraphics[height= 7.5
cm,width=8.0cm]{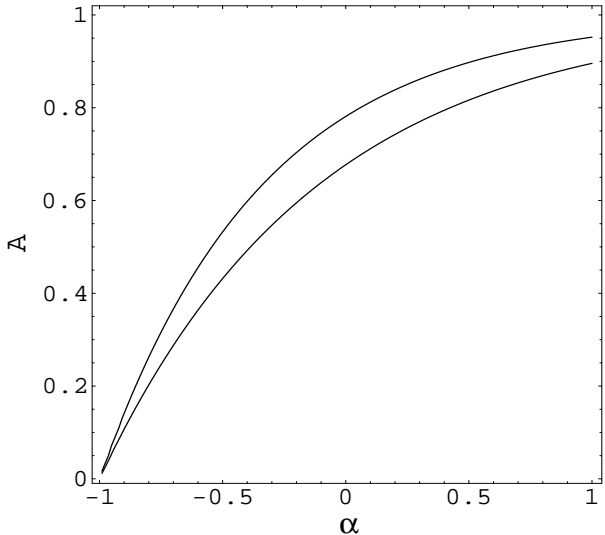} \caption{Contours of constant
$\Omega_{m}^{\text{eff}}$, corresponding to
$\Omega_{m}^{\text{eff}}=0.25$ and $\Omega_{m}^{\text{eff}}=0.35$.}%
\end{figure}

Independent constraints on Chaplygin models from galaxy clusters
x-ray data have been presented recently in Ref. \cite{cunha03}. In
that work, different constraints were obtained; for instance, the
contours are quite insensitive to the parameter $A$. In fact,
those authors were interested in quintessence Chaplygin, that is,
models in which the Chaplygin component behaves only as dark
energy, while here we are considering quartessence Chaplygin (no
dark matter). Therefore, since Ref. \cite{cunha03} started from
different priors and did not consider that the Chaplygin component
can cluster, it is natural that it reached different conclusions.

It is worth noting that the contours that we have obtained with
the Chandra data correspond to constraints on
$\Omega_{m}^{\text{eff}}$. To illustrate this, we plot in Fig. $5$
the lines $\Omega_{m}^{\text{eff}}=0.25$ and
$\Omega_{m}^{\text{eff}}=0.35$. The similarity of the contours is
evident. We remark that constraints on the QCM models similar to
those obtained here should be expected from large-scale structure
data (assuming entropy perturbations as in \cite{reis03}) that are
very sensitive to $\Gamma _{\text{eff}}$, the QCM effective shape
parameter of the mass power spectrum \cite{sugiyama95,beca03},
\begin{equation}
\Gamma_{\text{eff}}=\Omega_{m}^{\text{eff}}h\exp\left(  -\Omega_{b}%
-\frac{\sqrt{2h}\Omega_{b}}{\Omega_{m}^{\text{eff}}}\right)  \,.
\end{equation}

\section{Combined analysis and Discussion\label{disc}}

Here we summarize our results, presenting a combined analysis of
the constraints discussed in the previous sections. In Fig. $6$,
we display the allowed region of the parameters $A$ and $\alpha$
from a combination of data from FR IIb galaxies and gravitational
lensing. Although $A$ is better constrained than in each one of
these experiments separately, it is still not possible to
constrain $\alpha$ within our expected interval. Even if we
include type Ia supernovae data, as in Fig. $7$, we are not able
to constrain this parameter (although the best-fit value is
already inside the expected interval); only $A$ can be fairly well
constrained with this set. However, the inclusion of cluster x-ray
gas mass fraction data, combined with the previous three, as in
Fig. $8$, places significant constraints on both $A$ and $\alpha
$. These are the tightest constraints on the QCM parameters set to
date from the background geometry. The allowed interval of
$\alpha$, $-0.5\lesssim \alpha\lesssim1$, falls within the
expected interval discussed in Sec.~\ref{4essence}. Intriguingly,
the best-fit value of $\alpha$ is near the $\Lambda$CDM limit of
the model, $\alpha=0$.

It is seen that QCM models with $\alpha\gtrsim-0.5$ are consistent
with the observables considered here. Moreover, even the Chaplygin
($\alpha=1$) model---the most theoretically motivated of the QCM
family---cannot be ruled out by current observational data.

The aim of this paper was to verify whether the quartessence Chaplygin model
is consistent with current data on the background geometry, and to set
constraints on the model parameters. As we have discussed in Sec.\ref{intro},
even if the $\Lambda$CDM model is in agreement with the data, it is still
interesting to look for alternative models, both from a philosophical point of
view and also to test the robustness of the model against observational data,
i.e., to sense to which extent we can modify the $\Lambda$CDM paradigm and
still be in agreement with the data.

\begin{figure}[ptb]
\centering\hspace*{-0.8in} \includegraphics[height= 7.5
cm,width=8.0cm]{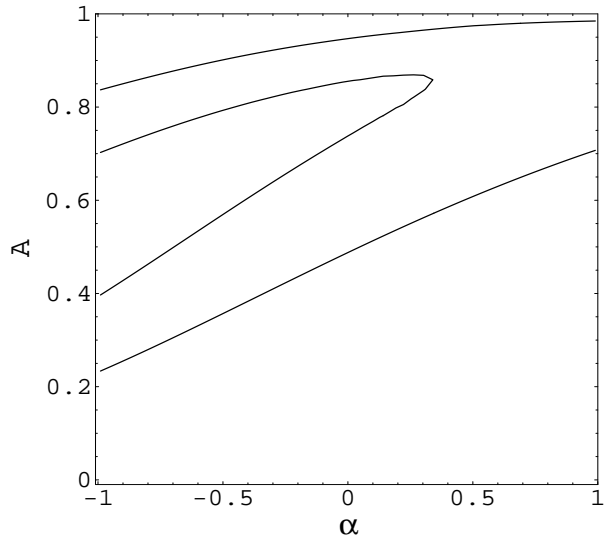}
\caption{Combined results from the
statistics of gravitational lensing and radio galaxies data. We
display the $68.3\%$ and $95.4\%$ confidence levels in
the ($\alpha,A$) plane.}%
\end{figure}

\begin{figure}[htb]
\centering\hspace*{-0.8in} \includegraphics[height= 7.5
cm,width=8.0cm]{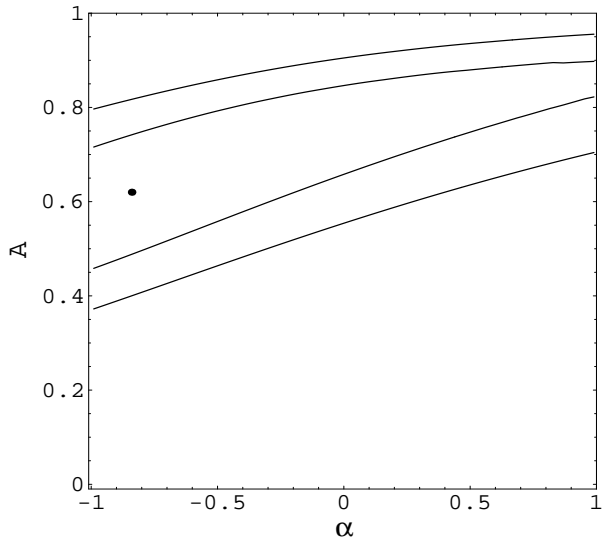} \caption{Combined results from supernovae,
gravitational lensing statistics, and radio galaxies data. We
display the $68.3\%$ and $95.4\%$ confidence levels in the
($\alpha,A$) plane. The dot is the best-fitting value for the
combination of these experiments.}%
\end{figure}

\begin{figure}[htb]
\centering\hspace*{-0.8in} \includegraphics[height= 7.5
cm,width=8.0cm]{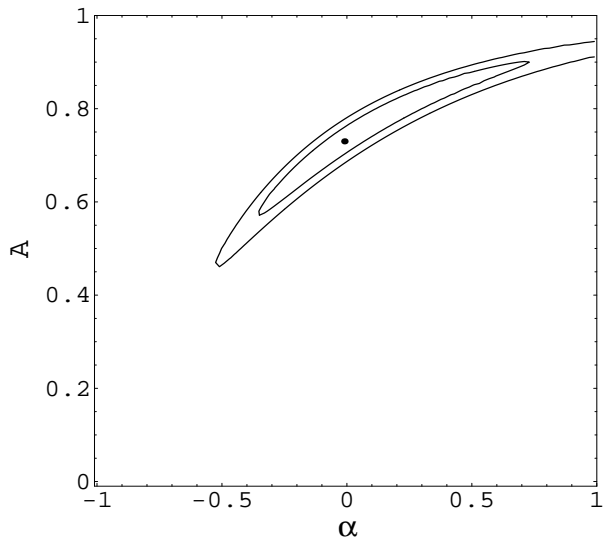} \caption{Confidence levels ($68.3\%$ and
$95.4\%$) from the combination of all the observables considered
in this paper (supernovae, gravitational lensing statistics, radio
galaxies, and cluster x-ray gas mass fraction). The best-fit
value is located close to the $\Lambda$CDM limit of the model.}%
\end{figure}

We argue that, as a model for dark energy (explanation for accelerated
expansion), the Chaplygin gas is not particularly attractive. The most
promising feature of it is in the context of quartessence (unification of dark
matter and dark energy). Here we have focused on the realization of the
quartessence scenario with a fluid whose background equation of state is given
by Eq. (\ref{eostoy}). Rather than favoring a specific model, our results show
that alternatives to $\Lambda$CDM are consistent with the data.

We have addressed only observables that probe essentially the
distance-redshift relation, and it is clear that one should look
for independent tests, such as, for instance, in the large-scale
structure of the Universe. In fact, the QCM has a rich behavior
regarding the density perturbations. For adiabatic perturbations,
the Chaplygin power spectrum has strong oscillations
\cite{sandvik02}. However, for positive values of $\alpha $, the
baryon power spectrum is well behaved \cite{beca03}. For some
entropy perturbations, the QCM power spectrum itself is well
behaved, even for negative values of $\alpha$ \cite{reis03}.
Therefore, one should investigate further the large-scale
distribution of the dark component to better constrain the QCM. In
particular, the nonlinear regime of structure formation in QCG
models should be more studied. This should shed some light on the
issue of the separation into clustered CDM-like and low-density
negative pressure regions. The impact on the CMB must also be more
studied \cite{carturan02,bean03,bento03,amendola03}.

Of course our analysis could be improved in several aspects. For instance, in
the lensing analysis, radio sources could be included. Also, we fixed some
parameters to their best-fitting values, and they could be marginalized to
provide more robust constraints. However, we expect that this will not change
qualitatively the results presented here.

The QCM still seems to be a promising model for unifying dark
matter and dark energy. More generically, the idea of quartessence
has to be explored further, both from the particle physics point
of view---to search for a first principles motivation to it---as
well as from the empirical side, to constrain quartessence models
from observational data.

\bigskip\textbf{Acknowledgments}

M.M. is partially supported by the FAPERJ. S.Q.O. is partially supported by
CNPq. I.W. is partially supported by the Brazilian research agencies CNPq and
FAPERJ. We would like to thank Ruth Daly and Erick Guerra for helpful
discussions and Steve Allen for useful suggestions and for sending us the gas
mass fraction data. We wish also to acknowledge Maur\'{i}cio O. Calv\~{a}o for
useful comments and discussions. M.M. acknowledges the hospitality of the
University of Arizona, where part of this work was done.

\end{document}